\documentclass[epj]{svjour}

\usepackage{bm}
\usepackage{amssymb}
\usepackage{color}
\usepackage{mathtools}
\usepackage{tikz}
\usepackage{url}
\usepackage{hyperref}
\usepackage{cite}
\usetikzlibrary{decorations.pathmorphing,decorations.markings}
\usetikzlibrary{decorations.pathreplacing}
\usetikzlibrary{positioning, calc, shapes}
\usetikzlibrary{shapes.geometric, arrows}
    \tikzstyle{startstop} = [rectangle, rounded corners, minimum width=2cm,text centered, draw=black, fill=red!30]
    \tikzstyle{check} = [rectangle, minimum width=2.0cm, text centered, draw=black, fill=blue!30]
    \tikzstyle{iout} = [trapezium, trapezium left angle=-70, trapezium right angle=-110, minimum width=3cm, text centered, draw=black, fill=blue!30]
    \tikzstyle{io} = [trapezium, trapezium left angle=70, trapezium right angle=110, minimum width=3cm, minimum height=1cm, text centered, draw=black, fill=blue!30]
    \tikzstyle{nnpdf} = [rectangle, minimum width=3cm, minimum height=1cm, text centered, draw=black, fill=orange!30]
    \tikzstyle{n3py} = [rectangle, minimum width=3cm, minimum height=1cm, text centered, draw=black, fill=green!30]
    \tikzstyle{n3cpp} = [rectangle, minimum width=3cm, minimum height=1cm, text centered, draw=black, fill=blue!30]
    \tikzstyle{procblue} = [rectangle, minimum width=3cm, minimum height=1cm, text centered, draw=black, fill=blue!30]
    \tikzstyle{fitted} = [rectangle, minimum width=5cm, minimum height=1cm, text centered, draw=black, fill=red!30]
    \tikzstyle{fixed} = [rectangle, minimum width=5cm, minimum height=1cm, text centered, draw=black, fill=blue!30]
    \tikzstyle{arrow} = [thick,->,>=stealth]

\definecolor{darkgreen}{rgb}{0.0, 0.5, 0.13}
\usepackage{xcolor}

\begin{document}

\title{Towards a new generation of parton densities\\ with deep learning models}

\author{Stefano Carrazza \and Juan Cruz-Martinez}
%
%
\institute{TIF Lab, Dipartimento di Fisica, Universit\`a degli Studi di Milano and
        INFN Sezione di Milano,\\ Via Celoria 16, 20133, Milano, Italy}

\date{Received: date / Revised version: date}
\abstract{We present a new regression model for the determination of parton
  distribution functions (PDF) using techniques inspired from deep learning projects.
  In the context of the NNPDF methodology, we implement a new efficient computing
  framework based on graph generated models for PDF parametrization and gradient
  descent optimization. The best model configuration is derived from a robust
  cross-validation mechanism through a hyperparametrization tune procedure. We
  show that results provided by this new framework outperforms the current
  state-of-the-art PDF fitting methodology in terms of best model selection and
  computational resources usage.
} 
\maketitle

\section{Introduction}

In perturbative QCD, parton distribution functions (PDFs) are used to describe the non-perturbative structure of hadrons~\cite{Butterworth:2015oua,Kovarik:2019xvh,AbdulKhalek:2019mps}.
These functions are typically determined by means of a supervised regression model which compares a wide set of experimental data with theoretical predictions computed with a PDF parametrization.
A truthful determination of PDFs and its uncertainties are important requirements when producing theoretical prediction for precision studies in high energy physics.
From a methodological point of view, the choice of a regression model and its
uncertainty treatment is a crucial decision which will impact the quality of
PDFs and its theoretical predictions.

The aim of this paper is to describe a new regression strategy framework inspired on deep
learning techniques for the NNPDF
methodology~\cite{Ball:2017nwa}. The NNPDF methodology uses
machine learning techniques in combination of Monte Carlo data generation to
extract PDFs from experimental data. The NNPDF approach was pioneer in
using artificial neural networks for the PDF parametrization and genetic
algorithms for optimization.
The NNPDF fitting framework has been constantly reviewed and upgraded in the last years, by including new features and methodological improvements which enhanced the quality of the released PDF sets~\cite{Bertone:2017tyb,Carrazza:2017udv,Bertone:2018ecm}.
Motivated by the new technologies and algorithms in use by the machine learning community, we dedicate this study to asses the impact of such new strategies in a modern PDF determination.

We focus our study on three issues. The first consists in improving performance
of the current NNPDF approach, where each PDF replica fit requires a large
number of CPU hours to complete, \textit{e.g.} in a global PDF determination a single fit takes
$\mathcal{O}(30)$ CPU hours. The second regards the efficiency (or lack thereof)
of neural network optimization through genetic algorithms.
Finally we aim to achieve a flexible framework in order to easily
change neural network models and tune its architecture and learning strategy.

The paper is organized as follows. In Section~\ref{sec:methodology}
we summarize briefly the current NNPDF methodology, highlighting the main differences with respect to the new approach
proposed in this paper as well as the testing setup we use for benchmarking
and tuning the results. In Section~\ref{sec:hyperopt}, we introduce the hyperparametrization procedure adopted to find the optimal learning strategy.
Finally, in Section~\ref{sec:results} we show as example some preliminary fits using this new technology.

\section{Methodology}
\label{sec:methodology}

\subsection{The NNPDF methodology}
\label{sec:nnpdfmethodology}

The NNPDF collaboration implements by default the
Monte Carlo approach to PDF fits. The goal of such strategy is to provide an
unbiased determination of PDFs with reliable uncertainty. The NNPDF methodology
is based on the Monte Carlo treatment of experimental data, the parametrization
of PDFs with artificial neural networks, and the minimization strategy based on
genetic algorithms.

In the next paragraphs we outline the most relevant aspects of the NNPDF3.1
methodology. An exhaustive overview is beyond the scope of this paper, we
invite the reader to review~\cite{Ball:2017nwa,Ball:2014uwa} for further details.

The Monte Carlo approach to experimental data consists in generating artificial
data replicas based on the experimental covariance matrix of each experiment.
This procedure allows to propagate experimental uncertainties into the PDF
model by performing a PDF fit for each data replica. Usually, PDF sets generated
from such approach are composed by 100 to 1000 replicas.

The experimental data used in the PDF fit is preprocessed according to a cross-validation
strategy based on randomly splitting the data for each replica into a training set and a validation set.
The optimization is then performed on the training set while the
validation loss is monitored and used as stopping criterion to reduce overlearning.

In the NNPDF fits, PDFs are parameterized at a reference scale $Q_0$
and expressed in terms of a set of neural networks corresponding to a set of basis
functions. Each of these neural networks consists of a fixed-size feedforward multi-layer
perceptron with architecture 1-2-5-3-1.
The input node ($x$) is split by the first layer on the pair $(x, \log(x))$.
The two hidden layers (of 5 and 3 nodes) use the sigmoid activation function while the output node
is linear:
\begin{equation}
    f_i(x,Q_0) = A_i x^{-\alpha_i} (1-x)^{\beta_i} {\rm NN}_i(x), \label{eq:PDFdefinition}
\end{equation}
where ${\rm NN}_{i}$ is the neural network corresponding to a given flavour $i$,
usually expressed in terms of the PDF evolution basis
$\{g,\,\Sigma,\,V,\,V_3,\,V_8,\,T_3,\,T_8,\,c^+\}$. $A_i$ is an overall
normalization constant which enforces sum rules and $x^{-\alpha_i}
(1-x)^{\beta_i}$ is a preprocessing factor which controls the PDF behaviour at
small and large $x$. In order to guarantee unbiased results, in the current NNPDF
methodology both the $\alpha_i$ and $\beta_i$ parameters are randomly selected
within a defined range for each replica at the beginning of the fit and kept
constant thereafter.

Unlike usual regression problems, where during the optimization procedure
the model is compared directly to the training input data, in PDF fits the
theoretical predictions are constructed through the convolution operation per
data point between a fastkernel table (FK) as presented in Ref.~\cite{Ball:2010de,Bertone:2016lga},
which encodes the theoretical
computation, and the PDF model, following the process
type of the data point. For DIS-like processes the convolution is performed
once, while for hadronic-like processes PDFs are convoluted twice.

The optimization procedure consists in minimizing the loss function
\begin{equation}
    \chi^2 = \sum_{i,j}^{N_{\rm dat}} (D-P)_i \sigma_{ij}^{-1} (D-P)_j, \label{eq:chi2}
\end{equation}
where $D_i$ is the $i$-nth artificial data point from the training set, $P_i$ is the
convolution product between the fastkernel tables for point $i$ and the PDF
model, and $\sigma_{ij}$ is the covariance matrix between data points $i$
and $j$ following the $t_0$ prescription defined in appendix of Ref.~\cite{Ball:2012wy}. This covariance matrix can include both experimental and theoretical components as presented in Ref.~\cite{AbdulKhalek:2019bux}.

Concerning the optimization procedure, so far, only genetic algorithms (GA) have
been tuned and used.
In summary the  procedure consists in initializing the weights
of the neural network for each PDF flavour using a random gaussian distribution and
checking that sum rules are satisfied. From that first network 80 mutant copies are
created based on a mutation probability and size to update the weights.
The training procedure is fixed to 30k iterations
and stopping is determined using a simple look-back algorithm which stores the
best weights for the lowest validation loss value.

\subsection{A new methodological approach}

The methodology presented above is currently implemented in a C++ code, introduced for the
first time in official releases in NNPDF3.0~\cite{Ball:2014uwa} and which
relies on a very small set of external libraries.
This feature can become a shortcoming as the monolithic structure of the codebase greatly
complicates the study of novel architectures and the introduction of modern machine
learning techniques developed during the last few years.

Our target in this work is to construct a new framework that to allow for the
enhancement of the methodology. In order to achieve our goal we rebuild the code using
an object oriented approach that will allow us to modify and study each bit of the
methodology separately.

We implement the NNPDF regression model from scratch in a python-based framework in which every piece aims to be completely independent.
We choose Keras~\cite{chollet2015keras} and Tensorflow~\cite{tensorflow2015:whitepaper} in order to provide the neural network capabilities for the framework as they are some of the most used and well documented libraries, sometimes also used in the context of PDFs~\cite{Wang:2018heo,AbdulKhalek:2019mzd}.
In addition, the code design abstracts any dependence on these libraries in order to be able to easily implement other machine learning oriented technologies in the future.
This new framework, by making every piece subjected to change, opens the door to a plethora of new studies which were out of reach before.

For all fits shown in this paper we utilize gradient descent (GD) methods to substitute the previously used genetic algorithm. This change can be shown to greatly reduce the computing cost of a fit while maintaining a very similar (and in occasions improved) $\chi^{2}$-goodness. The less stochastic nature of GD methods also produces more stable fits than its GA counterparts. The main reason why the GD methods had not been tested before were due to the difficulty of computing the gradient of the loss function (mainly due to the convolution with the fastkernel tables) in a efficient way. This is one example on how the usage of new technologies can facilitate new studies thanks to differentiable programming and distributed computing.

We also use just one single densely connected network as opposed to a separate network for each flavour.
As previously done, we fix the first layer to split the input $x$ into the pair $(x, \log(x))$.
We also fix 8 output nodes (one per flavour) with linear activation functions.
Connecting all different PDFs we can directly study cross-correlation between the different PDFs not captured by the previous methodology.

As we change both the optimizer and the architecture of the network, it is not immediately obvious which would be the best choice of parameters for the NN (which are collectively known as hyperpameters). Thus, we implement in this framework the hyperopt library~\cite{Bergstra:2013} which allow us to systematically scan over many different combinations of hyperparameters finding the optimal configuration for the neural network. We detail the hyperparameter scan in Section~\ref{sec:hyperopt}.

\subsection{A new fitting framework: \texttt{n3fit}}

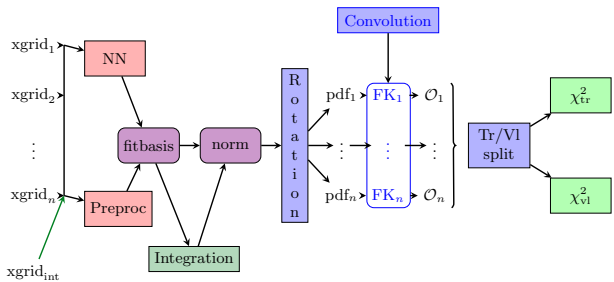
\begin{figure}[tb]
    \centering
    \resizebox{0.45\textwidth}{!}{%
    \begin{tikzpicture}[node distance = 1.0cm] \small
        \node (x1) {$\text{xgrid}_{1}$};
        \node[below of = x1] (x2) {$\text{xgrid}_{2}$};
        \node[below of = x2] (xd) {\vdots};
        \node[below of = xd] (xn) {$\text{xgrid}_{n}$};
        \node[below = 1cm of xn] (xint) {$\text{xgrid}_{\text{int}}$};

        \node[fitted, right = 1.0cm of x1.south, minimum width=1.2cm, minimum height=0.7cm]
                (pdf) {NN};
        \node[fitted, right = 1.0cm of xn.south, minimum width=1.2cm, minimum height=0.7cm]
                (preproc) {Preproc};

        \node[startstop, fill=violet!40, minimum width=1.2cm, minimum height=0.7cm, right = 1.5cm of xd.east]
                (fitbasis) {fitbasis};
        \node[startstop, fill=violet!40, minimum width=1.2cm, minimum height=0.7cm, right = 0.4cm of fitbasis]
                (normalizer) {norm};

        \node[fixed, minimum width=0.3cm, right = 0.4 of normalizer, text width=0.3cm]
                (rotation) {R o t a t i o n};

        \node[right = 0.5cm of rotation] (pd) {\vdots};
        \node[above of = pd] (p1){pdf$_{1}$};
        \node[below of = pd] (pn) {pdf$_{n}$};

        \node[blue,right = 0.6cm of pd] (fd) {\vdots};
        \node[blue,above of = fd] (f1) {FK$_{1}$};
        \node[blue,below of = fd] (fn) {FK$_{n}$};
        \node[procblue, minimum width = 2cm, minimum height=0.5cm, draw=blue, above = 1cm of f1] (convolution) {\color{blue}Convolution};
        \draw[draw=blue, rounded corners] (f1.north west) rectangle (fn.south east);

        \node[right = 0.6cm of fd] (od) {\vdots};
        \node[above of = od] (o1) {$\mathcal{O}_{1}$};
        \node[below of = od] (on) {$\mathcal{O}_{n}$};

        \node[fixed, right = 0.5cm of od, minimum width = 1.2cm, text width=1cm, minimum height=0.7cm]
                (trvl) {Tr/Vl split};
        \coordinate[right = 1cm of trvl] (cp);
        \node[n3py, above of = cp, minimum width = 1.2cm, minimum height=0.7cm]
                (chi2tr) {$\chi^{2}_\text{tr}$};
        \node[n3py, below of = cp, minimum width = 1.2cm, minimum height=0.7cm]
                (chi2vl) {$\chi^{2}_\text{vl}$};

            \node[procblue, fill=darkgreen!30, minimum width=1.0cm, minimum height=0.5cm] at ($(preproc) + (1.5, -1.0)$)
                (integration) {Integration};

        \draw[arrow] (pdf) -- (fitbasis);
        \draw[arrow] (preproc) -- (fitbasis);
        \draw[arrow] (fitbasis) -- (normalizer);
        \draw[arrow] (normalizer) -- (rotation);
        \draw[arrow] (trvl) -- (chi2tr);
        \draw[arrow] (trvl) -- (chi2vl);
        \draw[arrow] (integration) -- (normalizer);
        \draw[arrow] (fitbasis) -- (integration);
        \draw[arrow] (rotation) -- (p1);
        \draw[arrow] (rotation) -- (pd);
        \draw[arrow] (rotation) -- (pn);
        \draw[arrow] (p1) -- (f1);
        \draw[arrow] (pd) -- ($(fd) + (-0.33, 0.0)$);
        \draw[arrow] (pn) -- (fn);
        \draw[arrow] (f1) -- (o1);
        \draw[arrow] ($(fd)+(0.33,0.0)$) -- (od);
        \draw[arrow] (fn) -- (on);
        \draw[arrow] (convolution) -- (f1.north);

        \draw[decorate, decoration={brace}, thick] (o1.north east) -- (on.south east);


        \coordinate (a1) at ($(x1) + (0.6, 0.0)$);
        \draw[arrow] (x1) -- (a1);
        \draw[arrow] let
                \p1 = (a1), \p2 = (x2) in
                (x2) -- (\x1, \y2);
        \draw[arrow] let
                \p1 = (a1), \p2 = (xn) in
                (xn) -- (\x1, \y2);

        \draw[thick] let
                \p1 = (a1), \p2 = (xn) in
                (a1) -- (\x1, \y2);

        \draw[arrow] (a1) -- (pdf);
        \draw[arrow] let
                \p1 = (a1), \p2 = (xn) in
                (\x1, \y2) -- (preproc);

        \draw[arrow, darkgreen] let
                \p1 = (a1), \p2 = (xn) in
                (xint) -- (\x1, \y2);
    \end{tikzpicture}
    }
    \caption{Diagrammatic view of the \texttt{n3fit} code. Each different block is set as a different layer, following a structure similar to Keras. The red squared blocks correspond to blocks with fittable parameters.}
    \label{fig:n3fit}
\end{figure}

In Fig.~\ref{fig:n3fit} we show a schematic view of the full new methodology which we will refer to from now on as \texttt{n3fit}.
The $\text{xgrid}_{1}\dots \text{xgrid}_{n}$ are vectors containing the $x$-inputs of the neural network for each of the experiments entering the fit.
These values of $x$ are used to compute both the value of the NN and the preprocessing factor, thus computing the unnormalized PDF.
The normalization values $A_{i}$ are then computed at every step of the fitting (using the $\text{xgrid}_{int}$ vector as input), updating the ``norm'' layer and producing the corresponding normalized PDF of Eq.~\eqref{eq:PDFdefinition}.

Before obtaining a physical PDF we apply a basis rotation from the fit basis, $\{g,\,\Sigma,\,V,\,V_3,\,V_8,\,T_3,\,T_8,\,c^+\}$, to the physical one, namely, $\{\bar{s}, \bar{u}, \bar{d}, g, d, u, s, c(\bar{c})\}$.
After this procedure we have everything necessary to compute the value of the PDF for any flavour at the reference scale $Q_{0}$.

All fittable parameters live in the two red blocks, the first named NN (by default a neural network composed by densely connected layers corresponding to the NN of Eq.~\eqref{eq:PDFdefinition}) and the second the preprocessing $\alpha$ and $\beta$ which are free to vary during the fit (in NNPDF3.1 for each replica $\alpha_{i}$ and $\beta_{i}$ are fixed during the fit).
In the next, when we refer to the neural network parameters we will be referring collectively to the parameters of these two blocks.

As in this new methodology each block is completely independent we can swap them at any point, allowing us to study how the different choices affect the quality of the fit. All the hyperparameters of the framework are also abstracted and exposed (crucial for the study shown in Section~\ref{sec:hyperopt}). It also allow us to study many different architectures unexplored until now in the field of PDF determination.

The PDFs, as seen in Section~\ref{sec:nnpdfmethodology}, cannot be compared directly to data, therefore it is necessary to bring the prediction of the network (the pdf$_{i}$ of Fig.~\ref{fig:n3fit}) to a state in which it can be compared to experimental data. For that we need to compute the convolution of the PDFs with the fastkernel tables discussed in Section \ref{sec:nnpdfmethodology} which produces a bunch of observables $\mathcal{O}_{1}\dots\mathcal{O}_{n}$ with which we can compute the loss function of Eq.~\eqref{eq:chi2}.

The first step of the convolution is to generate a rank-4 luminosity tensor (for DIS-like scenarios this tensor is equivalent to the PDF):
\begin{equation}
    \mathcal{L}_{i\alpha j \beta} = f_{i\alpha}f_{j\beta},
\end{equation}
where the latin letters refer to flavour index while the greek characters refer to the index on the respective grids on x.
The observable is then computed by contracting the luminosity tensor with the rank-5 fastkernel table for each separate dataset.
\begin{equation}
    \mathcal{O}^{n} = {\rm FK}^{n}_{i\alpha j \beta} \mathcal{L}_{i\alpha j \beta}, \label{eq:convolution}
\end{equation}
where $n$ corresponds to the index of the experimental data point within the dataset.
The computation of the observables is the most computationally expensive piece of the fit and the optimization and enhancement of this operation will be the object of future studies.

\begin{figure}[tb]
    \centering
    \begin{tikzpicture}[node distance = 1.2cm]
        \node (init) [startstop] {training step};
        \node (ccheck) [check, below of = init] {counter $>$ max};
        \node (pcheck) [check, below of = ccheck] {positivity $>$ threshold};
        \node (xcheck) [check, below of = pcheck] {$\chi^2_\text{val}$ $<$ best $\chi^2$};
        \node (reset) [startstop, text width = 3cm, below of = xcheck] {reset counter  best $\chi^2 = \chi^2_\text{val}$};Q

        \node (cplus) [startstop, left = 0.7cm of ccheck] {counter ++};
        \node (end) [startstop, right = 0.7cm of ccheck] {END};

        \coordinate[above of = cplus] (li);
        \coordinate[below of = cplus] (lp);
        \coordinate[below of = lp] (lx);
        \coordinate[below of = lx] (lr);

        \draw[arrow] (init) -- (ccheck);
        \draw[arrow] (ccheck) -- node[right] {No} (pcheck);
        \draw[arrow] (pcheck) -- node[right] {Yes}(xcheck);
        \draw[arrow] (xcheck) -- node[right] {Yes}(reset);

        \draw[arrow] (ccheck) --  node[above] {Yes} (end);

        \draw[arrow] (reset) -- (lr) -- (lx);
        \draw[arrow] (xcheck) -- node[above] {No} (lx) -- (lp);
        \draw[arrow] (pcheck) --  node[above] {No} (lp) -- (cplus);
        \draw[arrow] (cplus) -- (li) -- (init);
    \end{tikzpicture}
    \caption{Flowchart describing the patience algorithm of the \texttt{n3fit} code. The positivity constraint becomes more and more restrictive as the fit advances.}
    \label{fig:stopping}
\end{figure}
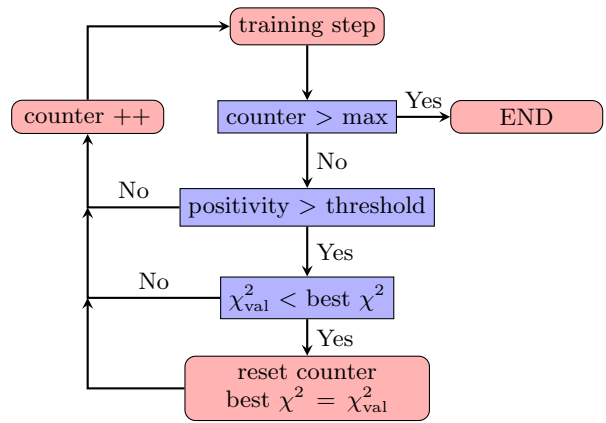

Before updating the parameters of the network we split the output into a training and validation set (selected randomly for each replica) and monitor the value of $\chi^{2}$ for each one of these sets. The training set is used for updating the parameters of the network while the validation set is only used for early stopping. The stopping algorithm is presented in Fig.~\ref{fig:stopping}. We then train the network until the validation stops improving. From that point onwards, and to avoid false positives, we enable a patience algorithm which waits for a number of iterations before actually considering the fit finished.

A last block to review is the positivity constraint in Fig.~\ref{fig:stopping}. We only accept points for stopping for which the PDF is known to produce positive predictions for special set of pseudo data which tests the predictions for multiple processes in different kinematic ranges $(x, Q^2)$. This mechanisms follows closely that used in previous versions of NNPDF~\cite{Ball:2014uwa,Ball:2017nwa}.

The loss function defined in Eq.~\eqref{eq:chi2} is minimized in order to obtain the best set of parameters for the NN. We restrict ourselves to the family of Gradient Descent algorithms with adaptive moment where the learning rate of the weights is dynamically modified.
In particular we focus on Adadelta~\cite{DBLP:adadelta}, Adam~\cite{Kingma:Adam} and RMSprop~\cite{Tieleman2012}. These three optimizer follow a similar gradient descent strategy but differ on the prescription for weight update.

\subsection{Environment setup: data and theory}
\label{sec:setup}
Benchmarking and validation of the new approach is done using as baseline the setup for NNPDF3.1 NNLO~\cite{Ball:2017nwa}. This means we will be using the same datasets and cuts, together with the same fraction of validation data for cross-validation although the stopping criteria is different (Fig.~\ref{fig:stopping}). This setup is named ``global'', as it includes all datasets used in NNPDF3.1 NNLO with 4285 data points.

We also define, in order to facilitate the process of benchmarking and validation, a reduced dataset with only DIS-type data with 3092 data points. Namely, all datasets from the ``global'' setup that are not hadronic. We call this setup ``DIS''. This reduced setup has a main advantage: in a DIS-like process there is only one PDF involved, this simplifies enormously the fit, making it much faster and lighter. These light fits, together with the new methodology, allow us to explore an space of parameters previously inaccessible.

\subsection{Performance benchmark}
\label{sec:performance}

\begin{table}[tb]
    \centering
    \begin{tabular}{c|ccc}
        \hline DIS fit & CPU h. & Mem. Usage (GB) & Good replicas \\ \hline
        \texttt{n3fit} (new)  & 0.2    & 2 & 95\%\\
        \texttt{nnfit} (old)  &  4    & 4 & 70\% \\
         \hline
    \end{tabular}
    \begin{tabular}{c|ccc}
        \hline Global fit & CPU h. & Mem. Usage (GB) & Good replicas \\ \hline
        \texttt{n3fit} (new)  & 1.5    & 14 & 95\%\\
        \texttt{nnfit} (old)  & 30     & 5 & 70\% \\
         \hline
    \end{tabular}
    \caption{Comparison of the average computing resources consumed by the old and new methodologies for the DIS and Global setups. We find \texttt{n3fit} to be $\sim20$ faster on average. The only drawback is the bigger memory consumption in the global fit. Each fit can be comprised of 100 to 200 replicas. Good replicas are those which pass all post-fit criteria defined in~\cite{Ball:2014uwa}.}
    \label{table:benchmark}
\end{table}

In order to obtain a good quality and reliable PDF model it is necessary to perform the fit for many artificial data replicas.
These are complex computation which require a great deal of CPU hours and memory consumption, therefore one of the goals of any new studies is to find a more efficient way of performing the PDF fits.
As previously stated, GD methods improve the stability of the fits, producing less ``bad replicas'', which need to be discarded, than theirs GA counterparts and this translates to a much smaller computing time.
In Table~\ref{table:benchmark} we find a factor of 20 improvement with respect to the old methodology and near to a factor of 1.5 in the percentage of accepted replicas for a global fit setup.

In the old code the memory usage is driven by the usage of APFEL~\cite{Bertone:2013vaa}, which does not depend on the set of experiments being used.
Instead, the memory consumption of the new code is driven by the Tensorflow optimization strategy which in the case of hadronic data requires the implementation of Eq.~\eqref{eq:convolution} and its gradient.
This difference translates to an importance increase on the memory usage of \texttt{n3fit} that is only realized in the Global fit.

We are currently working on ways that would allow us to reduce the memory consumption without introducing a penalty on the execution speed of the code as currently we favour speed with respect to memory.

\section{Hyperparameter tuning}
\label{sec:hyperopt}

The NNPDF approach aimed to reduce the bias introduced in the determination of the functional form utilized to parametrize the PDFs~\cite{Forte:2002fg}. Neural networks provide universal function approximators~\cite{Cybenko1989} which reduce systematic biases introduced by the choice of specific functional forms. Neural networks themselves, however, are not unique and the space of hyperparameters is big enough that finding the best choice becomes a overwhelming task.

In this work we aim to improve over the previous iteration of the NNPDF methodology by boxing the entire framework under hyperparameter optimization routines, there are several key points which allow us to do this now. Firstly, the new design of the code exposes all parameters of the fit including (but not restricted to) the neural network architecture. This is of key importance for a proper hyperparameter scan where everything is potentially interconnected. Secondly, the new methodology has such a smaller impact on computing resources that we can perform more fits on a scale of orders of magnitude, in other words, for each fit using the old methodology we can now test hundreds of architectures.

The hyperparameter scan capabilities are implemented using the hyperopt framework~\cite{Bergstra:2013} which systematically scans over a selection of parameter using Bayesian optimization~\cite{Bergstra:2011:AHO:2986459.2986743} and measures model performance to select the best architecture.

As a proof of concept, for this paper we make a first selection of parameters on which to scan, shown in Table~\ref{table:scanparameters} separated between the parameters which define the Neural Network architecture and those which define the fitting procedure.

\begin{table}
    \centering
    \begin{tabular}{c c} \hline
        Neural Network & Fit options \\ \hline
        Number of layers (*) & Optimizer (*) \\
        Size of each layer & Initial learning rate (*) \\
        Dropout & Maximum number of epochs (*) \\
        Activation functions (*) & Stopping  Patience (*) \\
        Initialization functions (*) & Positivity multiplier (*) \\
        \hline
    \end{tabular}
    \caption{Parameters on which the hyperparameter scan is performed. Results marked with (*) are shown graphically in Fig.~\ref{fig:hyperoptDIS}.}
    \label{table:scanparameters}
\end{table}

In this study we apply the framework to both the global and DIS only setup and in order to achieve the best model configuration we limit the data input to the experimental central values instead of using artificial replicas. We optimize on a combination of the best validation loss and stability of the fits. In other words, we select the architecture which produces the lowest validation loss after we trim those combinations which are deemed to be unstable.

\begin{figure*}
    \center
    \includegraphics[width=1.0\textwidth]{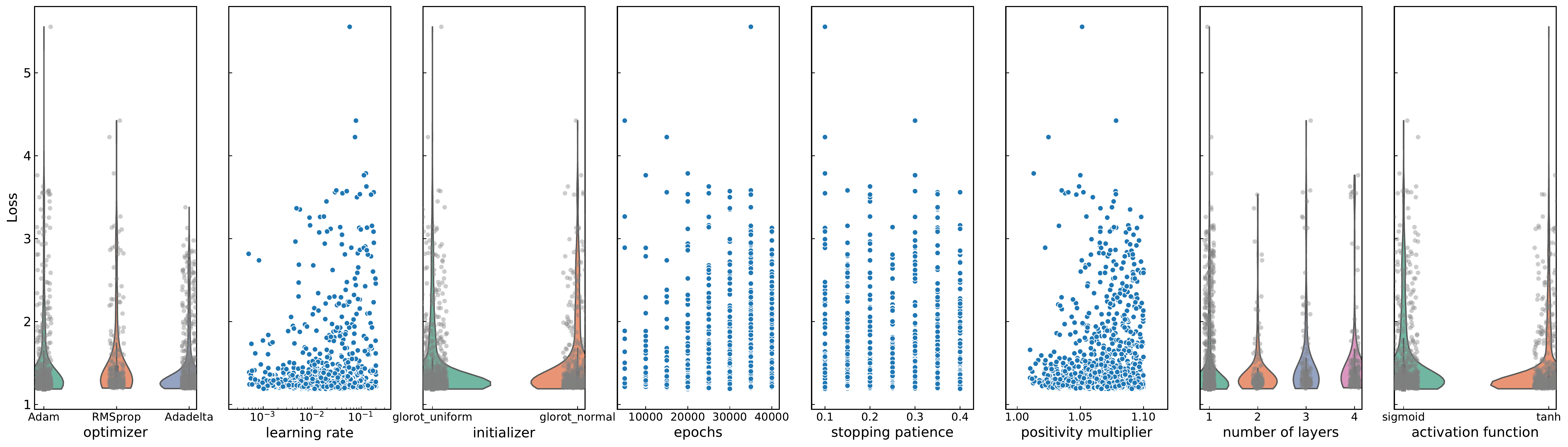}
    \caption{Graphical representation of an hyperparameter scan for a DIS only fit with 2000 trials. The loss presented in the y-axis corresponds to an average of the validation and testing loss functions. The shape of the violin plots represent a visual aid on the behaviour of the fit as a function of the free parameter. Fatter plots represent better stability, {\it i.e.}, configurations which are less likely to produce outliers.}
    \label{fig:hyperoptDIS}
\end{figure*}

In Fig.~\ref{fig:hyperoptDIS} we show an example of DIS only scan. We present eight examples of those shown in Table~\ref{table:scanparameters}

In this scan we find the Adadelta optimizer, for which no learning rate is used, to be more stable and systematically produce better results than RMSprop and Adam with a wide choice of learning rates.
The initializers, once unstable options such as a random uniform initialization have been removed, seem to provide similar qualities with a slight preference for the ``glorot\_normal'' initialization procedure described in~\cite{Glorot10understandingthe}.

Concerning the parameters related to the stopping criteria, we observe that when the number of epochs is very small the fit can be certainly unstable, however after a certain threshold no big differences are observed. The stopping patience shows a very similar pattern, stopping too early can be disadvantageous but stopping to late does seem to make a big difference. The positivity multiplier, however, shows a clear preference for bigger numbers.

Finally, concerning the neural network architecture we observe that a small number of layers seem to produce slightly better absolute results, however, one single hidden layer seem to be also very inconsistent. The activation functions present with a slight preference for the hyperbolic tangent. Once we have a acceptable hyperparameter setup we ran again for fine tuning as some of the choices could have been biased by a bad combination on the other parameters.

The main take away from this scan is the implementation of a semi-automatic methodology able to find the best hyperparameter combination as the setup changes, \textit{e.g.} with new experimental data, new algorithms or technologies.

\subsection{Overfitting: the test set}

While performing the hyperparameter scan we found that optimizing only looking at the validation loss produced results which would usually be considered overfitted: very low training and validation $\chi^2$ and complex replica patterns.
Thanks to the high performance of the \texttt{n3fit} procedure the usual cross-validation algorithm used in the NNPDF framework was not enough to prevent overlearning for all architectures.

The cross-validation implemented in NNPDF is successful on avoiding the learning of the noise within a dataset.
However, we observe that this choice is not enough to prevent overfitting due to correlations within points in a same dataset when using hyperopt with {\tt n3fit}.
In order to eliminate architectures that allowed overlearning we proceed by including a testing set where the model generalization power is tested. This is a set of datasets where none of the points are used in the fitting either for training or validation.

Defining the best appropriate test dataset for PDF fits is particularly challenging due to the nature of the model regression through convolutions.
For the present results the test set is defined by removing from the training set datasets with duplicate process type and smaller leading-order kinematic range coverage.
We call the loss produced by the removed datasets ``testing loss'' and we use it as a third criterion (beyond stability and combined with the value of the validation loss) to discard combinations of hyperparameters.
With this procedure we are able to find combinations of hyperparameters which produce good fits for which we are confident no obvious overfitting can be generated.
In Table~\ref{table:testset} we list the datasets which have been used as test set for this study.

\begin{table}[tb]
    \centering
    \begin{tabular}{lc} \hline
        Experiment & Observable \\ \hline
        NMC & $\sigma_{\rm NC}^e$ \\
        BCDMS & $F^p_2$, $F^d_2$ \\
        HERA & $\sigma_{\rm NC}^p$, $F_2^b$ \\
        D0 & $\frac{d \sigma_{Z}}{d y_z}$, W electron asy\\
        CDF & $k_{t}$ incl jets\\
        ATLAS & mass DY, $\sigma (t\bar{t})$\\
        CMS & Z $p_{T}$ 8 TeV \\
        \hline
    \end{tabular}
    \caption{List of datasets used for the testing procedure of the hyperparameter scan. After each fit the generalization power of the network is tested on these sets and the iteration discarded if its $\chi^{2}$ greatly differs from the validation and training $\chi^{2}$. This list is a subset of the datasets entering the fits for NNPDF 3.1~\cite{Ball:2017nwa}.}
    \label{table:testset}
\end{table}

\begin{figure}[tb]
    \includegraphics[width=0.5\textwidth]{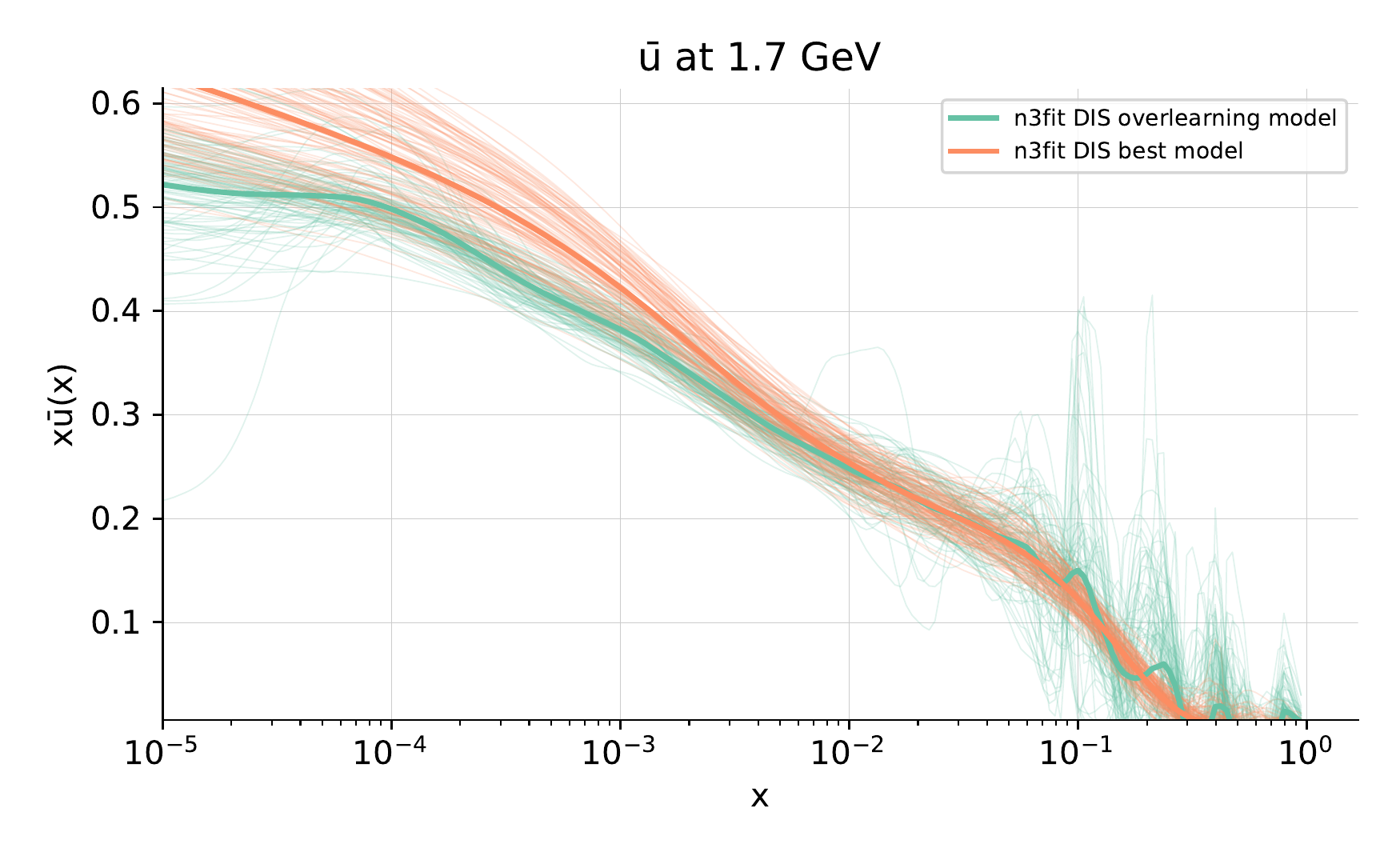}
    \caption{Comparison between the PDF replicas generated by \texttt{n3fit} for one parton flavour ($\bar{u}$). In green we show the results for the best model in the naive hyperoptimization, in orange the best model once we have introduced the test set criteria.}
    \label{fig:overfitted}
\end{figure}

In Fig.~\ref{fig:overfitted} we show an example of a PDF produced by two very different architectures, both of which are generated by the hyperoptimization procedure. We observe a much more unstable behaviour in the fit in which we allow for overtraining which in turn translates for a $\chi^{2}$ on the testing set of more than twice the value of the training set. We believe the issue of hidden correlations in experimental measurements as well as its impact on PDF fits requires a much deeper study outside the scope of this paper.

\section{Results}
\label{sec:results}

\begin{table}[tb]
    \centering
    \begin{tabular}{l|cc}
        \hline Parameter  & DIS only & Global \\ \hline
         Hidden layers & 2 & 3 \\
         Architecture & 35-25-8 & 50-35-25-8 \\
         Activation & tanh & sigmoid \\
         Initializer & glorot\_normal & glorot\_normal \\
         Dropout & 0.0 & 0.006 \\
         Optimizer & Adadelta & Adadelta \\
         Max epochs & 40000 & 50000 \\
         Stopping patience & 30\% & 30 \% \\
         \hline
    \end{tabular}
    \caption{Best models found by our hyparparameter scan for the DIS and global setups using the new \texttt{n3fit} methodology.  }
    \label{table:bestModel}
\end{table}

The best setups we find are shown in Table~\ref{table:bestModel}. We find the global setup allow for deeper networks without falling in overfitting.
The hyperbolic tangent and the sigmoid functions are found to perform similarly.
The initializer of the weights of the network, however, carries some importance for the stability of the fits.
We utilize the glorot normal initialization method~\cite{Glorot10understandingthe,glorot:2010} as implemented in Keras.

We find that adding a small dropout rate~\cite{DBLP:journals/corr/abs-1207-0580} to the hidden layers in the global fit reduces the chance of overlearning introduced by the deeper network, thus achieving more stable results.
As expected the bigger network shows a certain preference for greater waiting times (which also increases the stopping patience as is set to be a \% of the maximum number of epochs). In reality the max. number of epochs is rarely reached and very few replicas are wasted.

\begin{table}[tb]
    \centering
    \begin{tabular}{c|cc}
        \hline  & DIS only & Global \\ \hline
        \texttt{n3fit} (new) & 1.10 & 1.15 \\
        \texttt{nnfit} (old) & 1.13 & 1.16  \\
         \hline
    \end{tabular}
    \caption{Comparison of the total $\chi^{2}$ of the fit for both a DIS only and Global setup between the old and new methodology.}
    \label{table:chi2fits}
\end{table}

We have produced two complete fits for the DIS only and global setups described in Sec.~\ref{sec:setup}.
We find that both fits perform similarly on describing the experimental data, as can be attested by the values of $\chi^{2}$ presented in Table~\ref{table:chi2fits}. Below we detail the results separating between the DIS only and global setups and showing a direct comparison between the behaviour of the PDFs found with both methodologies.
In all plots of this section the orange color corresponds to the fit performed with the old methodology whereas green corresponds to the new one and are generated using \texttt{reportengine}~\cite{zahari_kassabov_2019_2571601}.

\subsection{DIS only fits}

\begin{figure}[tb]
    \includegraphics[width=0.5\textwidth]{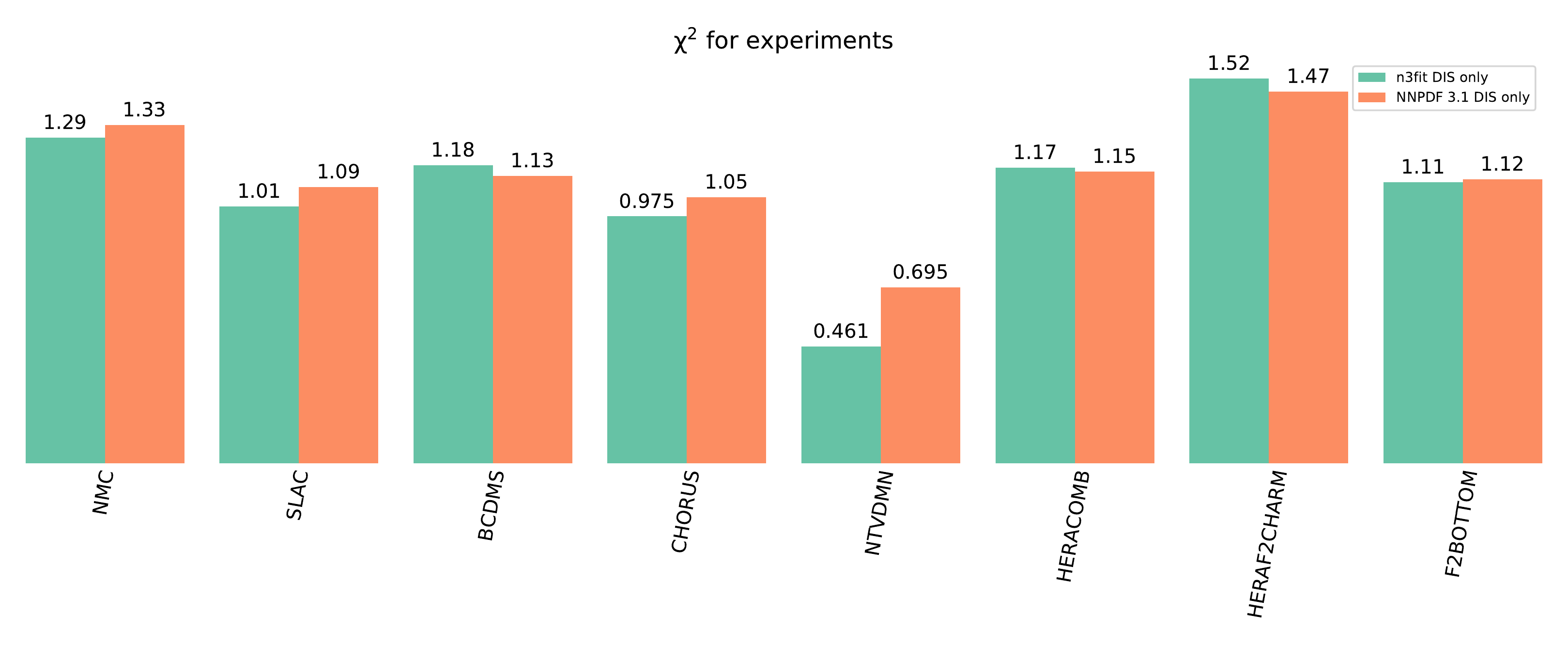}
    \caption{Comparison of the $\chi^{2}$ experiment by experiment between the new and old codes. An experiment is comprised of one or more datasets of experimental data. All datasets in this plot correspond to DIS-type experiments.}
    \label{fig:DISExChi2}
\end{figure}
\begin{figure}[tb]
    \includegraphics[width=0.5\textwidth]{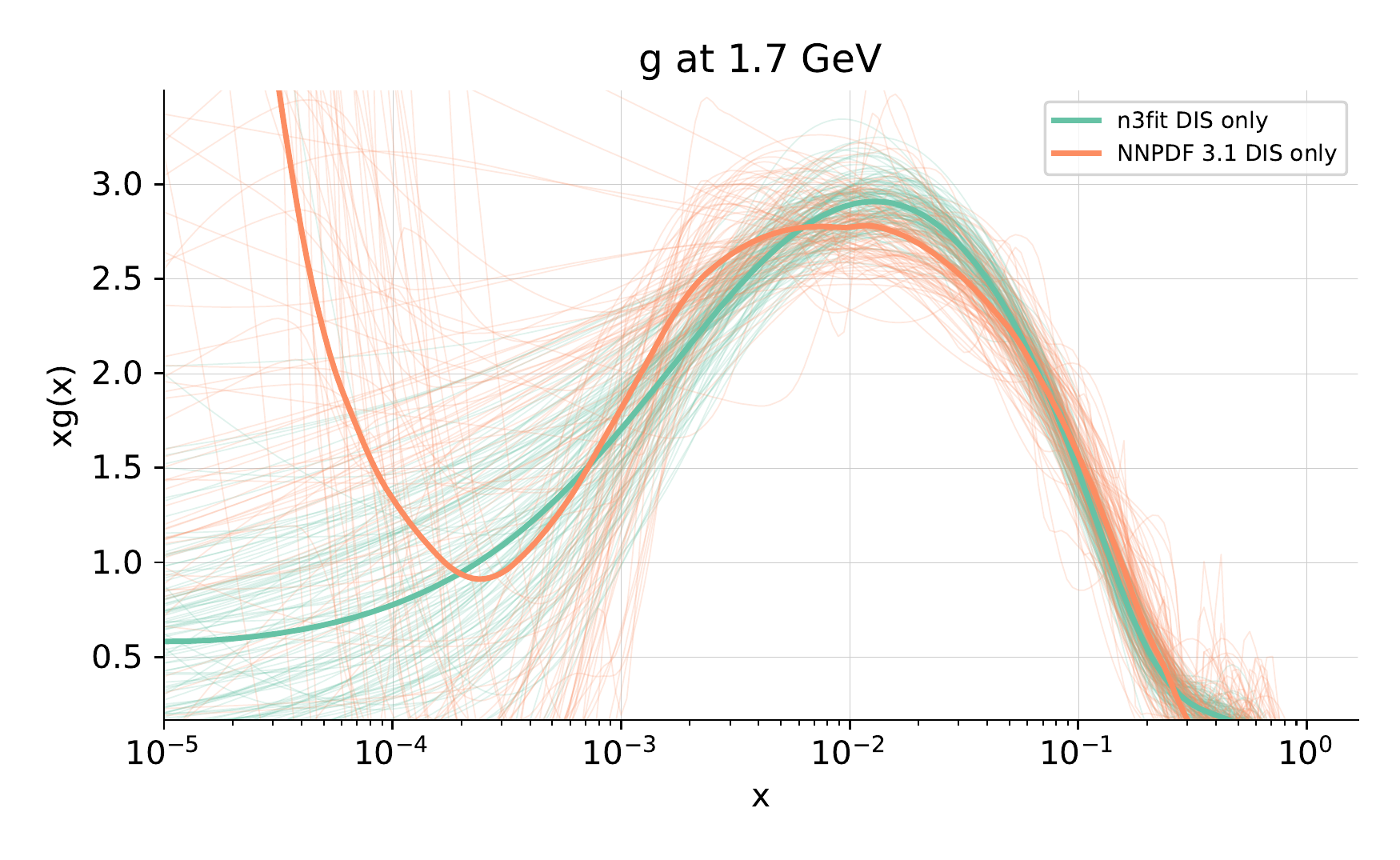}
    \includegraphics[width=0.5\textwidth]{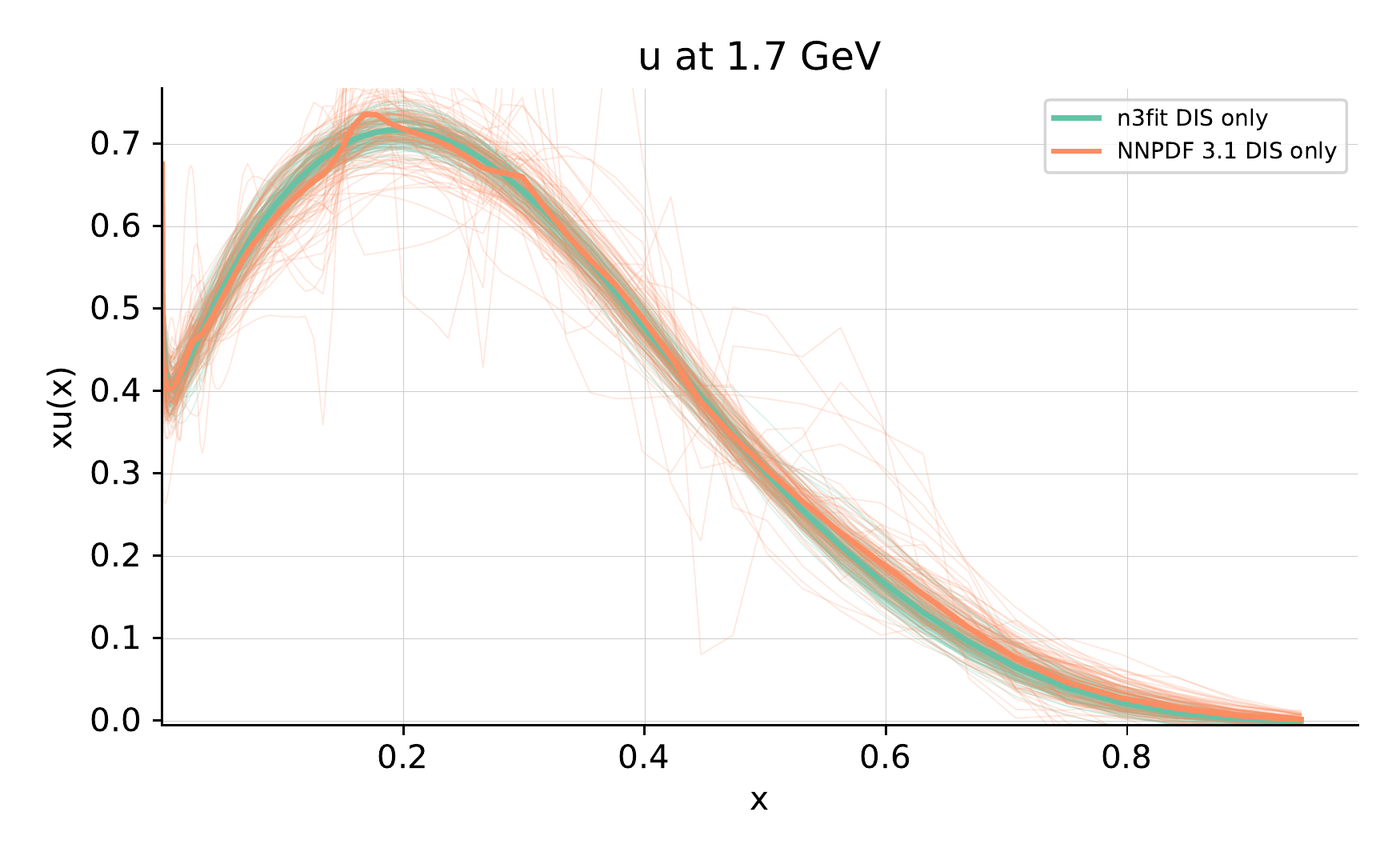}
    \caption{Comparison of the PDF of the gluon and the $u$-quark for the DIS only setup. Each line correspond to a different replica while the bold-faced line correspond to the central value of the PDF computed by taking the average of all other replicas.}
    \label{fig:DISpdf}
\end{figure}

When we study the change on the value of $\chi^{2}$ in Table~\ref{table:chi2fits} it is interesting to study also how this value changes experiment by experiment.
In Fig.~\ref{fig:DISExChi2} we compare the individual $\chi^2$ experiment by experiment between the new and old methodology.
We observe that values are compatible within the statistical fluctuations obtained by changing the random seed during the initialization of the old methodology.
From this we can infer the behaviour of the PDFs must also be similar. We show some examples for the gluon and $u$-quark replicas in Fig.~\ref{fig:DISpdf}. Indeed, the central value for the PDF is not very different from the one obtained with the old methodology (albeit somewhat smoother) and always lying within the one sigma band of the old PDF fits.

\begin{figure}[tb]
    \includegraphics[width=0.5\textwidth]{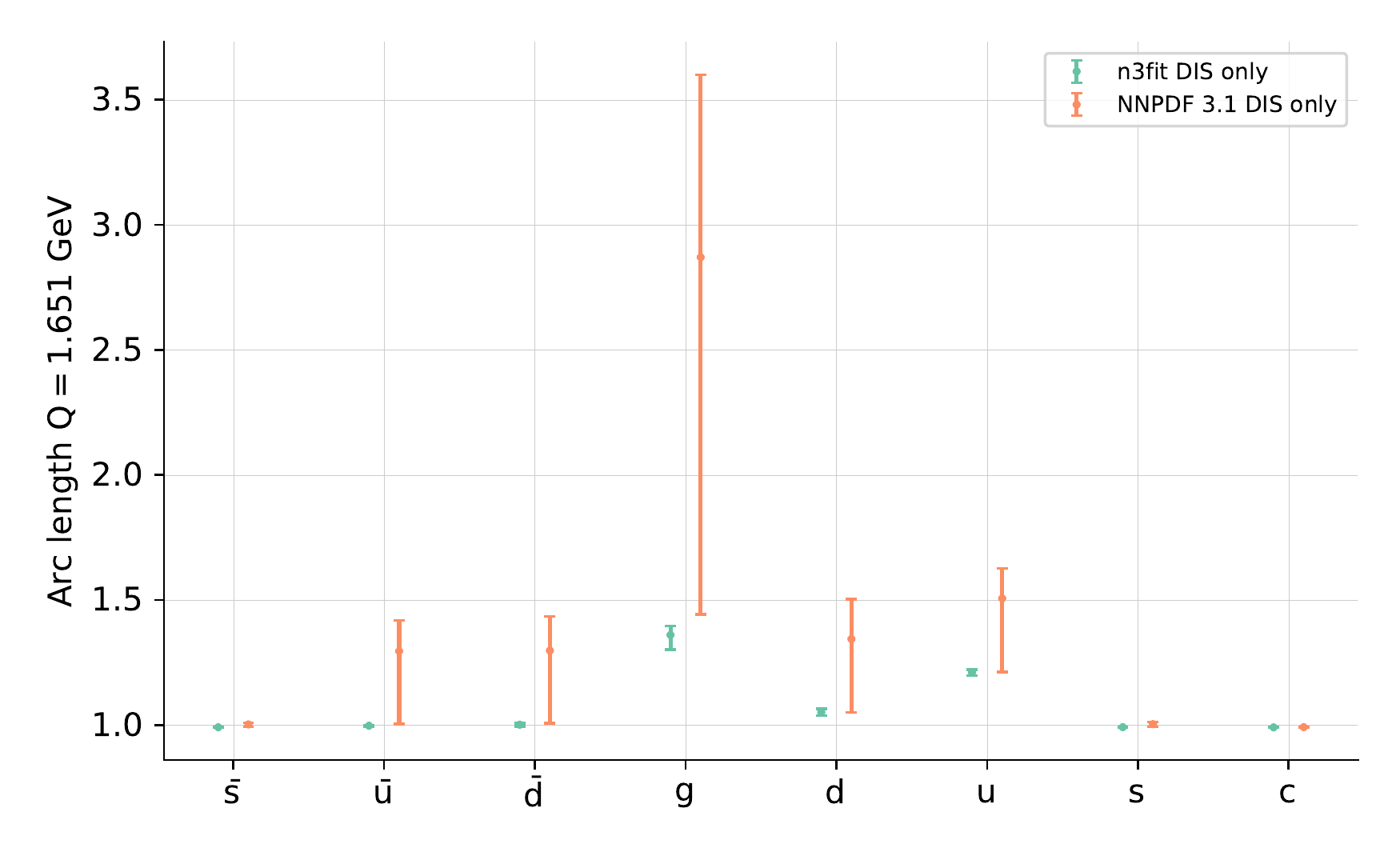}
    \caption{Comparison for the DIS only setup of the arc-lengths between the new and old methodologies. The new methodology produces smoother curves which translates to smaller and more stable arc-lengths.}
    \label{fig:DISArcLenghts}
\end{figure}

The biggest difference between both methodologies resides on the stability of the replicas.
In Fig.~\ref{fig:DISpdf} we can see that \texttt{n3fit} produces smoother replicas with less complex structure.
The central value, however, remains stable and well within the envelope of NNPDF3.1.
As a result in Fig.~\ref{fig:DISArcLenghts} we observe that the arc-lengths for all flavours are systematically smaller with {\tt n3fit}.

\subsection{Global fits}
\begin{figure}[tb]
    \includegraphics[width=0.5\textwidth, height=0.6\textheight]{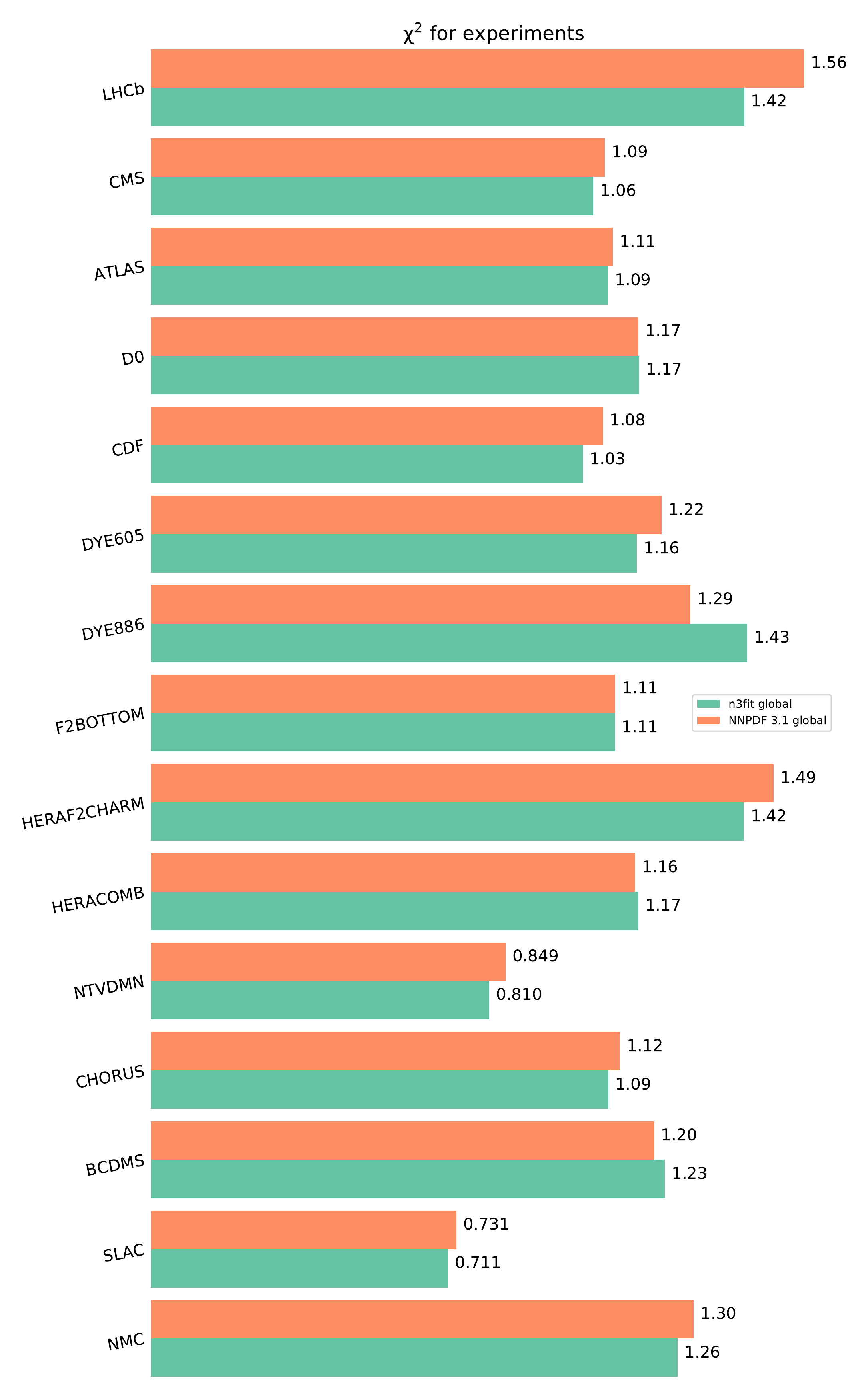}
    \caption{Comparison of the $\chi^{2}$ between both the new and old codes experiment by experiment for a global fit. An experiment is comprised of one or more datasets of experimental data. We find the new methodology to be able to produce fits with a quality similar to the old methodology for every experiment.}
    \label{fig:GlobalExChi2}
\end{figure}

\begin{figure}[tb]
    \includegraphics[width=0.5\textwidth]{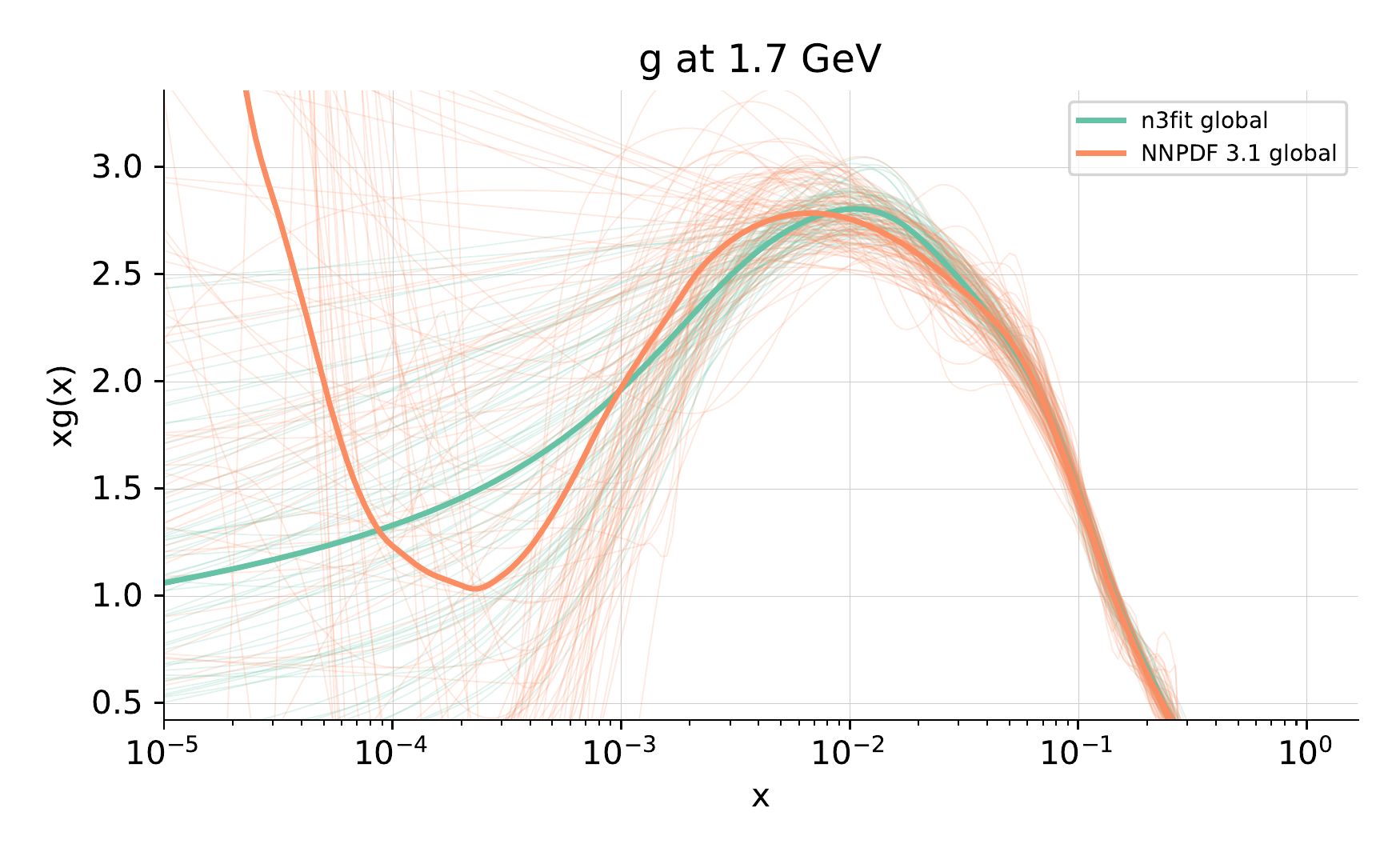}
    \includegraphics[width=0.5\textwidth]{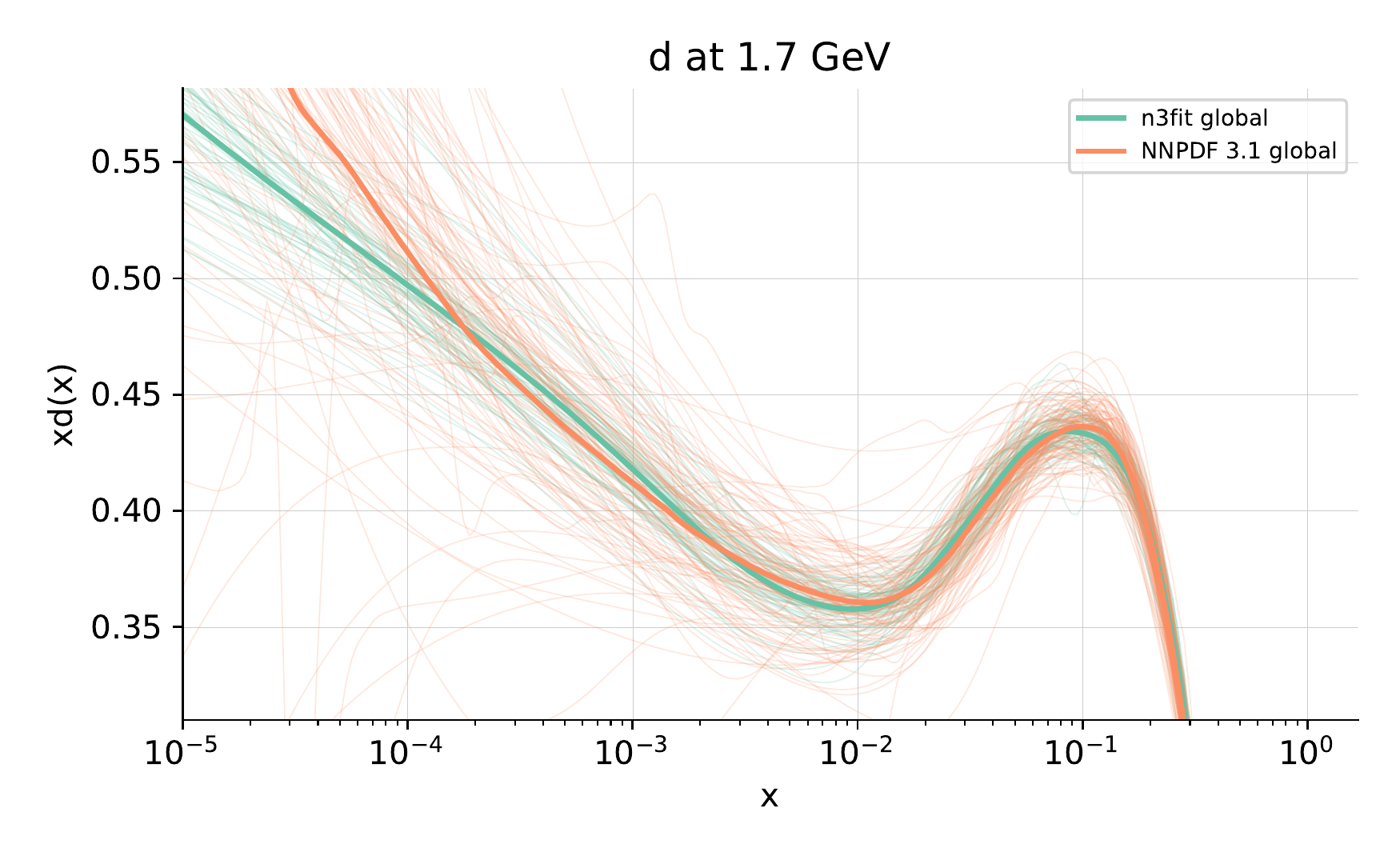}
    \caption{Comparison of the PDF of the gluon and the $d$-quark for the global setup. Each line correspond to a different replica while the bold-faced line correspond to the central value of the PDF computed by taking the average of all other replicas.}
    \label{fig:Globalpdf}
\end{figure}

Similar results are obtained in the case of the global fit. The $\chi^2$ per experiment is shown in Fig.~\ref{fig:GlobalExChi2}. Although the difference between the old and new methodology in the global setup is not as evident as in the DIS only case, we can still observe more stable replicas and in general smoother curves in Fig.~\ref{fig:Globalpdf}, where we plot all produced replicas for the new and old methodologies for the gluon and $d$-quark PDFs.

\begin{figure}[tb]
    \includegraphics[width=0.5\textwidth]{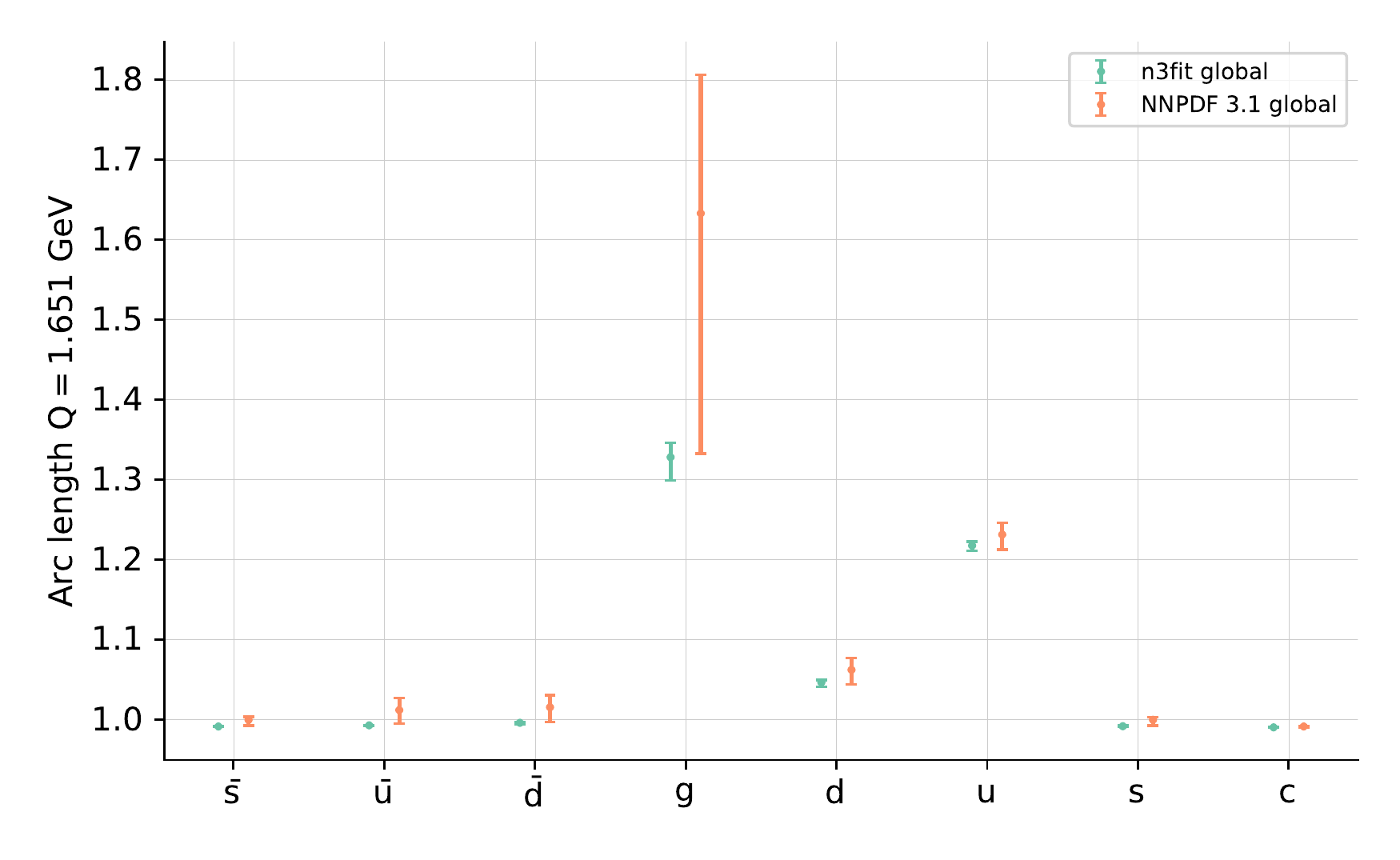}
    \caption{Comparison for the global setup of the arc-lengths between the new and old methodologies. As in the DIS setup, the new methodology produces smoother curves with smaller and more stable arc-lengths.}
    \label{fig:GlobalArcLenghts}
\end{figure}

The same is observed in Fig.~\ref{fig:GlobalArcLenghts} where the arc-lengths produced by \texttt{n3fit} are still systematically smaller.
It is also worth noticing the more stable behaviour of \texttt{n3fit} with respect to NNPDF3.1 when comparing the DIS only arc-lenghts of Fig.~\ref{fig:DISArcLenghts} with those produced by the global fit.

This difference can be easily understood as we now posses a framework able to scan and search for the best combination of hyperparameters for a given experimental setup which allow us to obtain good quality fits for any setup using the same framework.
Using the old methodology a similar study would have required several months of work.
This is another example of how this new methodology can improve the field of PDF determination.

\section{Conclusions}

We presented a new approach to and regression model strategy for the
determination of PDFs, in the context of the NNPDF framework. This new approach,
based on new computational techniques, improves fitting performance and quality.
Furthermore, we propose a new workflow pipeline for the systematic and
efficient determination of PDFs.

The new approach consists in replacing the current C++ fitting code with a new
implementation based on python and Keras-Tensorflow. This allows us to change model easily,
test new architectures developed by the scientific community, fit preprocessing
exponents, obtain faster results thanks to gradient based methods, and the
possibility to carry hyperparameter tuning in a systematic way to decide when
the model is optimal.

Finally, we believe that future PDF fits based on this new approach,
once exhaustively tested and validated,
will provide better and more reliable PDF sets for future releases of the NNPDF collaboration,
thanks to the possibility to identify in a quantitative way the best suited
regression model for new data from the LHC.

\section*{Acknowledgements}

We thank Stefano Forte, Juan Rojo and Jake Either for careful reading of the manuscript.
We also thank the members of the NNPDF collaboration for useful and insightful discussions.
This project is supported by the European Research Council under the European
Unions Horizon 2020 research and innovation Programme (grant agreement number
740006).

\bibliographystyle{epj}
\bibliography{../blbl}

\end{document}